\documentclass[aps,pre,groupedaddress,longbibliography]{revtex4-1}
\usepackage[utf8]{inputenc}
\usepackage{graphicx}
\usepackage{bm}
\usepackage{amsmath}
\usepackage{caption}
\usepackage{subfigure}
\usepackage{float}

\begin{document}
	\title{Generation and propagation of solitary waves in nematic liquid crystals}
	\author{Xingzhou Tang}
	\affiliation{Pritzker School of Molecular Engineering,
  University of Chicago, Chicago, Illinois 60637, U.S.A.}
	\author{Ali Mozaffari}
	\affiliation{Pritzker School of Molecular Engineering,
  University of Chicago, Chicago, Illinois 60637, U.S.A.}
	\author{Noe Atzin}
	\affiliation{Pritzker School of Molecular Engineering,
  University of Chicago, Chicago, Illinois 60637, U.S.A.}
	\author{Soumik Das}
	\affiliation{Smith School of Chemical and Biomolecular Engineering,
  Cornell University, Ithaca, New York 14853, U.S.A.}
	\author{Nicholas L. Abbott}
		\affiliation{Smith School of Chemical and Biomolecular Engineering,
  Cornell University, Ithaca, New York 14853, U.S.A.}
	\author{Juan J. de Pablo}
	\affiliation{Pritzker School of Molecular Engineering,
  University of Chicago, Chicago, Illinois 60637, U.S.A.}

	\begin{abstract}
		Solitons in nematic liquid crystals offer intriguing opportunities for transport and sensing in microfluidic systems. Little is known about the elementary conditions that are needed to create solitons in nematic materials. In this work, theory, simulations and experiments are used to study the generation and propagation of solitary waves (or "solitons") in nematic liquid crystals upon the application of an alternating current (AC) electric field. We find that these solitary waves exhibit "butterfly"-like or "bullet"-like structures that travel in the direction perpendicular to the applied electric field. Such structures propagate over long distances without losing their initial shape. The theoretical model adopted here serves to identify some of the key requirements that are needed to generate solitons in the absence of electrostatic interactions. These include surface imperfections that introduce a twist in the director, unequal elastic constants, and negative anisotropic dielectric permittivity. The results of simulations are shown to be in good agreement with our own experimental observations, serving to establish the validity of the theoretical concepts advanced in this work.
	\end{abstract}
	\maketitle
	
	\section{Introduction}
	
	Solitons are travelling wave packets that propagate at constant speed over long distances without losing their shape. A non-linear feedback mechanism during wave propagation serves to minimize their expansion (or dispersion)~\cite{Dauxois2006,Manton2004}. Solitons and solitary waves are abundant in nature; examples include water solitons in narrow channels, or the collective motion of proteins and DNA. They also arise in optical fibers, in magnets, and in nuclear physics~\cite{heimburg2005soliton,appali2012comparison,andersen2009towards,kartashov2019}. Several recent studies have reported the formation of solitons in liquid crystals (LCs) when subjected to an applied AC electric field~\cite{Lam1992}. Such studies have also outlined potential applications as long-distance information carriers for optical signals, or particles and chemical species within nematic LCs.
	
	Three-dimensional solitons can be generated in a nematic liquid crystal that exhibits a negative dielectric permittivity anisotropy upon application of an alternating electric field~\cite{Li2018,Aya2020}. Such solitons are usually observed in the vicinity of a surface imperfection, dust particles, or the edges of the electrodes, and they tend to travel over long distances in the direction perpendicular to the director field. Experiments indicate that, at low frequencies, solitons start as butterfly-like structures with quadrupolar symmetry. At higher applied fields and frequencies, the butterfly structure can emit bullet-like structures - solitons - having a curved director field. Such bullets can propagate at high speeds. In this work we introduce a model for the tensorial order parameter that is shown to be capable of predicting the formation of these solitary waves in good agreement with experimental observations.
	
	The results of simulations of such a model reveal that the key ingredients for soliton generation are (i) a negative anisotropy of permittivity, (ii) an irregularity that serves to nucleate solitons, and (iii) an AC field whose intensity and frequency must fall within a narrow range of values~\cite{das2022programming}. The results presented here are compared to experimental data for 4'butyl-4-heptyl-bicyclohexyl-4-carbonitrile $(CCN-47)$~\cite{Zawadzki2012,Lucchetti2021} - a liquid crystal for which detailed experimental observations have recently become available.

	\section{Theory and simulation}
	
	We consider a simple 3D thermotropic nematic liquid crystal. At each point in the material, the local orientational order is described in terms of the director field $\hat{\mathbf{n}}(\mathbf{r},t)$ or the nematic order tensor $Q_{ij}(\mathbf{r},t)$. The local fluid flow velocity is given by $\mathbf{v}(\mathbf{r},t)$. To solve the underlying model, we follow the method of Ref.~\cite{Tang2017, Tang2019} in 3 dimensions. The magnitude and the direction of the nematic order are allowed to vary, so that solitary waves and topological defects are able to form and move freely. We quantify the nematic order through a traceless, symmetric tensor of the form
	\begin{equation}
		Q_{ij} = s \left(\frac{3}{2}n_in_j-\frac{1}{2}\delta_{ij} \right).
	\end{equation}
	
	The free energy of the system can be expressed as
	\begin{align}
		F=\int d^2 r \biggl[&-\frac{1}{2}a|Q|^2+\frac{1}{3}b|Q|^3+\frac{1}{4}c|Q|^4\nonumber\\
		&+\frac{1}{2}\left(K_{11}- K_{24}\right)\left(Q_{il}\partial_jQ_{jl}\right)^2\\
		&+\frac{1}{2}\left(K_{22}- K_{24}\right)\left(\epsilon_{ijk}Q_{il}\partial_jQ_{kl}\right)^2\nonumber\\
		&+\frac{1}{2}K_{33}\left(Q_{il}\partial_iQ_{kl}\right)^2\nonumber+\frac{1}{2}K_{24}Tr\left(\bm{\Delta}^2\right)\nonumber\\
		&-\Delta\epsilon_0\epsilon_aE_i Q_{ij} E_j -\frac{4}{3}\chi_0E_i\partial_j Q_{ij}\nonumber\\
		&-\frac{4}{3}\chi_+E_i\partial_j\left(Q_{ik}Q_{jk}\right)-\frac{1}{3}\chi_2E_k\partial_k\left(Q_{ij}Q_{ij}\right)\nonumber\\
		&-\frac{4}{9}\chi_-E_i\left(Q_{ik}\partial_j Q_{jk}- Q_{jk}\partial_j Q_{ik}\right)\biggr].
		\label{freeenergy}
	\end{align}
	
	The first three terms, which consist of an expansion of the free energy in powers of $Q$, determine the magnitude of $\xi \sim \sqrt{c K / b^2}$ in the bulk, where gradients are small. The next four terms serve to quantify the Frank elastic free energy corresponding to splay, twist, bend and biaxial splay deformations of the material, respectively. In general, the full Oseen-Frank free energy density is written as ~\cite{selinger2018}
	\begin{align}
		f_{O-F}=&\frac{1}{2}K_{11}S^2+\frac{1}{2}K_{22}T^2+\frac{1}{2}K_{33}\bm{B}^2\nonumber\\
		&-\frac{1}{2}K_{24}\left[\frac{1}{2}S^2+\frac{1}{2}T^2-Tr\left(\bm{\Delta}^2\right)\right],
	\end{align}
where the last term is the saddle-splay term. Note that, in the literature, one finds several variations in the notation for the saddle-splay term ~\cite{selinger2018}. In particular, instead of $K_{24}$, the corresponding coefficient is sometimes written as $\frac{1}{2}K_{24}$ or as $\left(K_{22}+K_{24}\right)$. Those variations are not important for the arguments that follow. The often used approximation of equal elastic constants corresponds to $K_{11}= K_{22}= K_{33}=2 K_{24}\equiv K$.
	
	The subsequent terms in the free energy expression of Equation~(\ref{freeenergy}) correspond to the potential generated by the alternating current (AC) field. The first of these terms is the dielectric energy contribution, in which $\epsilon_0$ is the dielectric permittivity of vacuum and $\epsilon_a=\epsilon_\parallel-\epsilon_\perp$ is the permittivity anisotropy of the nematic material; $\epsilon_\parallel$ and $\epsilon_\perp$ are the dielectric permittivity in the directions perpendicular and parallel to the director, respectively. In materials with $\epsilon_a>0$, the liquid crystal aligns with the electric field, whereas in the negative case the alignment is transverse. A negative dielectric permittivity is required in our model, so that the director field favors alignment within the $x-y$ plane (the plane perpendicular to the electric field), as opposed to the $z-$axis (the direction of the electric field). The system experiences a sinusoidal electric field with frequency $\omega$ and maximum voltage $E_0$ along the $z-$axis, $\bm{E}=E_0 sin\left(\omega t\right)\hat{\bm{z}}$. 
	
	The terms with the $\chi$ coefficients represent the contributions of flexoelectricity ~\cite{Blow2013}. The $\chi_0$ and $\chi_+$ terms depend on spatial variations of the nematic order parameter and the electric field. Both terms are negligible due to the absence of spatial differences. The nematic order parameter, denoted by $S$, decreases at the location of topological defects or solitary waves. These regions are small (by a factor of about $10^3$) when compared to the entire sample size. The nematic order parameter can therefore be viewed as approximately uniform. The $\chi_2$ term vanishes when we calculate the relaxational dynamics by taking a partial derivative of the free energy. The $\chi_-$ term is a crucial component; the penalties for splay and bend deformations of the director field help amplify any irregularities created by surface imperfections, dust particles or inhomogeneities, which give rise to the nucleation of solitary waves. We note here that previous studies have also taken into account the $e_{11}$ and $e_{33}$ terms representing the effects of flexoelectricity in the director field~\cite{lavrentovich2020, alexe1993flexoelectric}, where the polarization is given by $\bm{P} = e_{11}\left(\hat{\mathbf{n}}\nabla\cdot\hat{\mathbf{n}}\right)-e_{33}\left(\nabla\times\hat{\mathbf{n}}\right)\times\hat{\mathbf{n}}$. Here $e_{11} = 2\chi_0+S\chi_+$, $e_{33} = S\chi_-$, where $S$ is the nematic order parameter.
	
	To model the time evolution, we use the following equations for dynamic relaxation
	\begin{align}
		\frac{\partial Q_{ij} (\bm{r},t)}{\partial t}=-\frac{1}{\gamma_1}\frac{\delta F}{\delta Q_{ij} (\bm{r},t)},
		\label{dynamicequations}
	\end{align}
	where $\gamma_1$ is the rotational viscosity. Hydrodynamic effects are not included in this first attempt to model nematic solitons.

	\section{Formation and propagation of solitons}
	
		To facilitate comparison of the predictions of our model to experiments we introduce several dimensionless quantities. In simulations, a characteristic length is defined by $\xi\sim \sqrt{cK/b^2}$, based on our free energy~(\ref{freeenergy}). A characteristic time $\tau$ is defined by the average time required by a bullet to propagate by one unit length, which amounts to $10^{6} s$ steps. In experiments, the characteristic length is the size of a bullet, approximately $20\sim50\mu m$. The characteristic time $\tau$ is the response time required by the nematic to return to equilibrium upon cessation of the electric field, $10^{-2} s$; that time is commensurate with the period of the AC field in our experiments. The electric field and frequency can be rewritten as $\widetilde{E} =\frac{\xi \chi_-}{K} E$ and $\widetilde{\omega}=\tau \omega$. 
		
		Both experiments and simulations reveal the existence of two types of solitary waves ~\cite{lavrentovich2020}. The first type consists of "butterfly" structures that, as already pointed out, tend to form around surface irregularities at low-to-intermediate applied fields and frequencies. The second consists of "bullet" structures, which are elongated entities that move rapidly and are observed at higher frequencies and voltages. Both structures bend along the direction of electric field ($z-$axis), and show an asymmetry in the direction of propagation of the solitons ($y-$direction). 
		
		%A symmetry can be figured out in the direction that most the director field aligning ($x-$direction). We suggest governing equations of bullets and butterflies based on a former research about the solitary structure Ref.~\cite{Calderer2019}.
	
	\begin{figure} [H]
		\centering
		\includegraphics[width=0.45\textwidth]{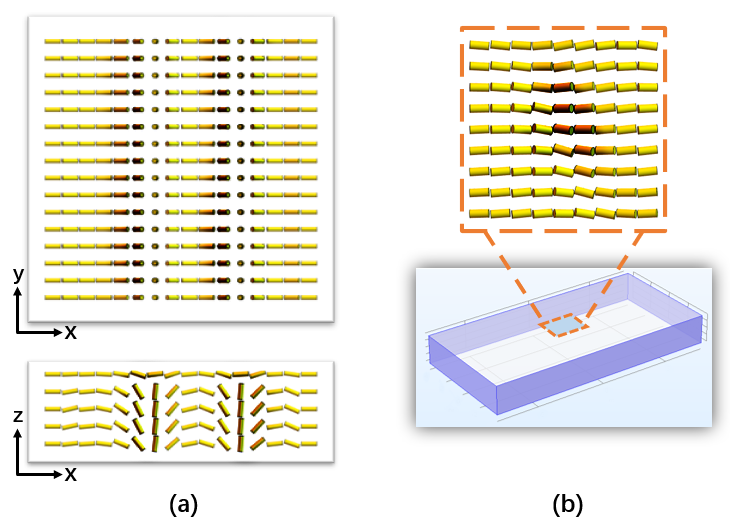}
		\includegraphics[width=0.45\textwidth]{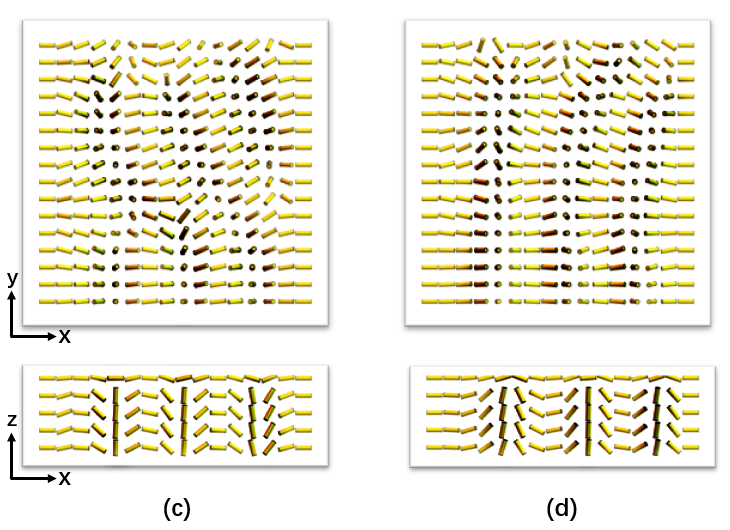}
		\caption{Director field induced by a DC field. The initial state, shown in (a), is completely uniform. (b) A small patch on the top surface is used to create a local tilt in the director in that region. (c) and (d) show the director patterns that arise above the surface patch as the direction of the electric field is alternated.}	
		\label{solitoncreat}
	\end{figure}
	
	The results of our simulations confirm that several conditions are necessary for solitons to emerge in a nematic liquid crystal. These are (1) a negative dielectric permittivity, (2) flexoelectricity, (3) a surface irregularity, (4) distinct Frank elastic constants for different deformation modes, and, (5) a periodic electric field. The dielectric permittivity, as mentioned the in last section, must be negative in order for the liquid crystal to align in the direction perpendicular to the electric field. At $45^{\circ}C$, the elastic constants for splay, bend and twist deformations of $CCN-47$ are $K_{11} = 7 pN$, $K_{22} = 1 pN$ and $K_{33} = 5 pN$ based on experimental measurements~\cite{das2022programming,lucchetti2021optical}. The $CCN-47$ liquid crystal offers a large, negative dielectric permittivity $(-6)$ at $45^{\circ}C$ Ref.~\cite{Dhara2008}. The dominant term of the flexoelectricity, $\chi_-$, is taken as $10^{-11} C/m$~\cite{Li2018, de1993physics}. FIG.~\ref{solitoncreat} shows the director pattern in the center of the sample ($x-y$ plane at $z=0$ and $x-z$ plane at $y=0$); one can appreciate the combined effects of flexoelectricity and surface irregularities. In the figure, we use rods to represent the 3D character of the director field. The color is light yellow when the director field aligns along the x-y plane, and it is dark when the director field aligns along the z-axis. The structures form in a short period of time after turning on the electric field. In FIG.~\ref{solitoncreat}a, the system is initially uniform. After applying a constant electric field, $\bm{E}=E_0 \hat{\bm{z}}$, the flexoelectricity acts to splay and bend the director field, and an asymmetry in the $z-$axis appears in the lower region of FIG.~\ref{solitoncreat}a. FIG.~\ref{solitoncreat} b shows that a small random tilt is induced in the center of the surface, which breaks the symmetry along the $y-$axis and generates a periodic pattern similar to that observed in experiments at low-frequency.  FIG.~\ref{solitoncreat} c and d show the same system as in b, but with the direction of the electric field reversed ($\bm{E}=E_0 \hat{\bm{z}}$ and $\bm{E}=-E_0 \hat{\bm{z}}$ respectively). The $x-z$ plane in FIGS.~\ref{solitoncreat} a and c are similar because the electric field is the same, but the structures are quite different when compared to c and d. In these two cases, a different direction of bending along the $z$-axis can be observed. One can appreciate that the patterns in the x-y plane are splayed and bent towards the opposite direction, but they are not complete mirror images of each other. A bias in the configuration appears, which is caused by the difference of the Frank constants. Under a one-constant assumption, that asymmetry would not arise.
	
	\begin{figure} [H]
		\centering
		\includegraphics[width=0.3\textwidth]{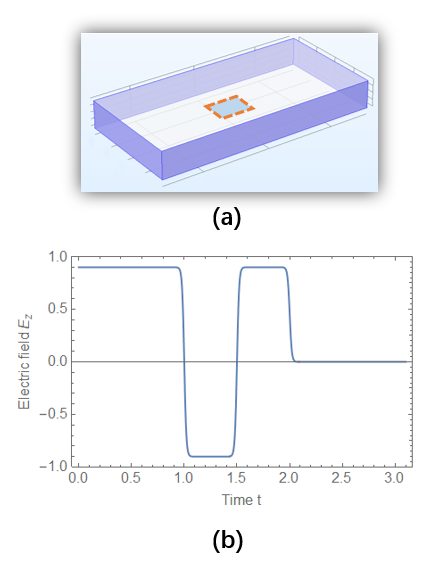}
		\includegraphics[width=0.45\textwidth]{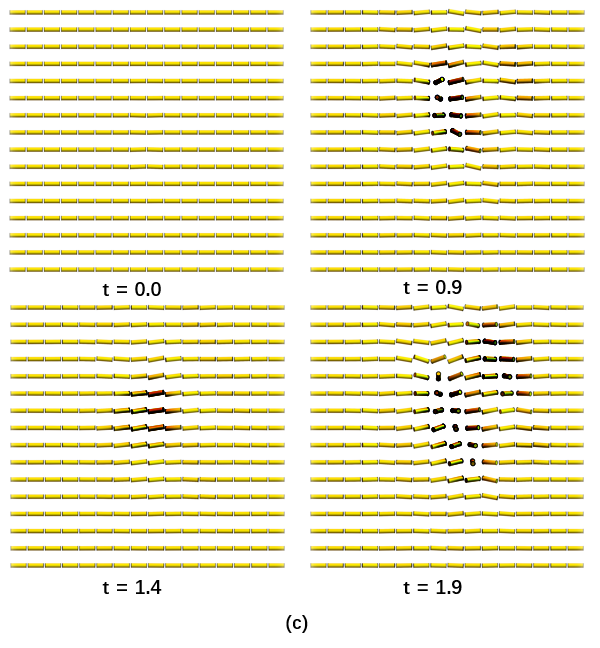}
		\caption{Director field induced by an electric field as it reverses direction. (a) Simulated system and (b) applied electric field. (c) Director field within the dashed area in (a) on the central x-y plane (z = 0) at different times.}
		\label{reverseDirection}
	\end{figure}
	
	FIG.~\ref{reverseDirection} shows the director pattern when the electric field is reversed. In this figure, the directors adopt a structure similar to that of a butterfly-like soliton as a result of the surface pattern and the flexoelectricity. This structure shrinks when the field is reduced, and it increases again when the electric field is raised again. The soliton propagates along the direction perpendicular to the alignment of the director field and the electric field. The flexoelectricity bends and splays the director field and breaks the symmetry in z, but the director field is mostly aligned along the plane perpendicular to the electric field ($x-y$ plane) due to the negative dielectric permittivity. The surface 
	irregularity breaks the symmetry along y, and creates a butterfly-like pattern around the patch. This structure can expand into a periodic pattern of the director field at locations far from the surface patch due to the effects of flexoelectric polarization. A slightly different pattern can be seen if the direction of the electric field is reversed because of the different Frank constants. As a result, an asymmetric structure occurs under the influence of the AC field.
	
	\begin{figure} [H]
		\centering
		\includegraphics[width=0.4\textwidth]{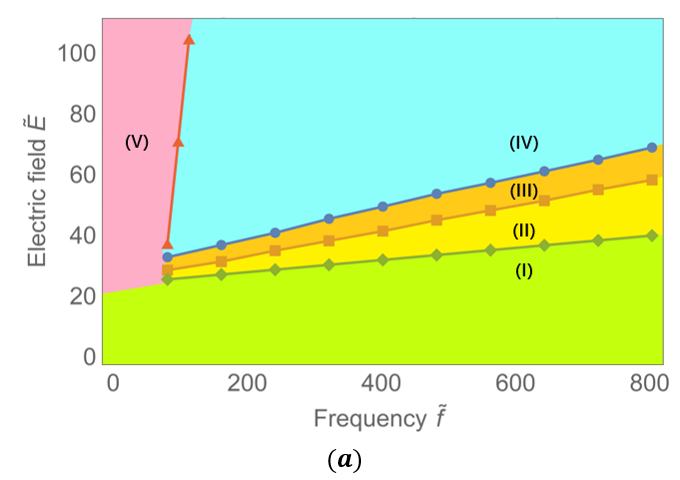}
	\end{figure}
	\begin{figure} [H]
		\centering
		\includegraphics[width=0.6\textwidth]{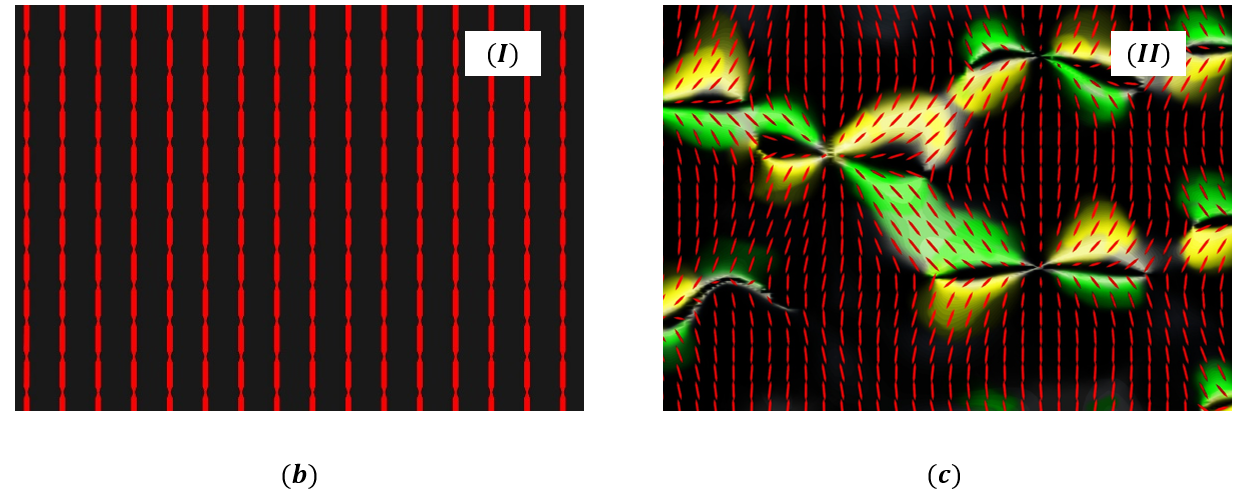}
	\end{figure}
	\begin{figure} [H]
		\centering
		\includegraphics[width=0.6\textwidth]{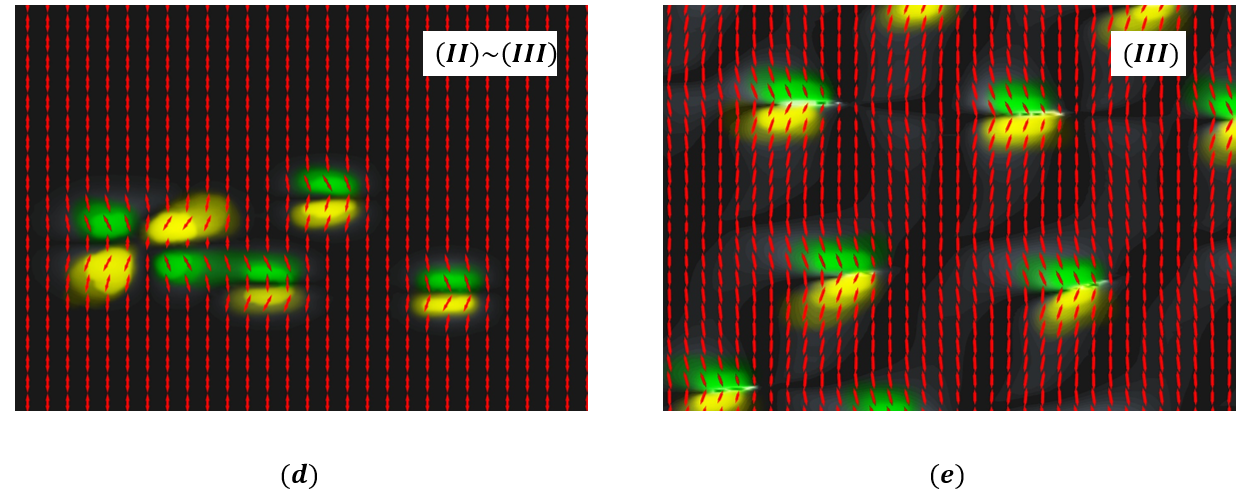}
	\end{figure}
	\begin{figure} [H]
		\centering
		\includegraphics[width=0.6\textwidth]{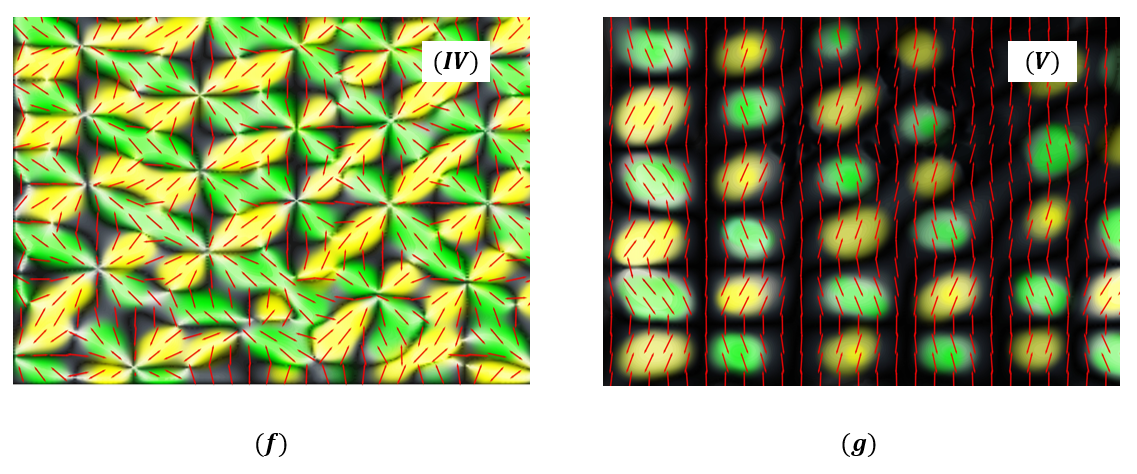}
		\caption{(a) State diagram showing the system's morphology as a function of voltage and frequency for a nematic liquid crystal under an applied AC field. (b)$\sim$(g) Starting from a uniform state (b), as the voltage is increased one observes the formation of "butterflies" in (c), followed by the formation of bullets emitted from the butterflies in (d). At higher voltages one enters a chaotic regime in (f). (g) represents the R-state that arises at low frequencies.}
		\label{phasediagram}
	\end{figure}
	
	In simulations we consider a channel with periodic boundaries in the x and y directions. The system size is $300\times200\times10 $. A butterfly-like irregularity is fixed on the top surface, and the director field everywhere else is initially aligned along the $x-$axis. A sinusoidal electric field $\bm{E}=E_0 sin \left(\omega t\right) \hat{\bm{z}}$ is applied, with voltage $ \widetilde{E_0} =0\sim100$, and frequency $ \widetilde{\omega} =0\sim800$. Different states can be observed for different values and frequency of the electric field. One can construct a phase diagram as a function of the electric field (FIG.~\ref{phasediagram}). Different states arise for different magnitudes of the frequency and field. We use red lines to represent the director field in a plane; the black color means the director field is uniform, and yellow and green represent different orientations when the director field is  aligned along the z-axis ~\cite{lavrentovich2020}. Section $\left(\uppercase\expandafter{\romannumeral1}\right)$ and $\left(\uppercase\expandafter{\romannumeral5}\right)$ correspond to cases with a small electric strength and a small frequency, respectively. In Section $\left(\uppercase\expandafter{\romannumeral1}\right)$, the field is too small to perturb the director field, which remains almost uniform. In Section $\left(\uppercase\expandafter{\romannumeral5}\right)$, the electric field is high enough to alter the director field, but the oscillation time of the field is longer than the response time of the material and a periodic pattern appears. This pattern is referred to as the R-state ~\cite{Aya2020} and it is similar to the structure observed under a DC field. Section $\left(\uppercase\expandafter{\romannumeral4}\right)$ shows that under high voltage, solitary waves are rarely observed, and the pattern is chaotic. In Section $\left(\uppercase\expandafter{\romannumeral2}\right)$ we observe butterfly-like structures, which upon further strengthening of the field start to emit bullets in the orange region between Section $\left(\uppercase\expandafter{\romannumeral2}\right)$ and Section $\left(\uppercase\expandafter{\romannumeral3}\right)$. When the voltage is increased even more, in Section $\left(\uppercase\expandafter{\romannumeral3}\right)$, a state with more bullets and stripes but fewer butterflies can be observed.
	
	\begin{figure} [H]
		\centering
		\includegraphics[width=0.55\textwidth]{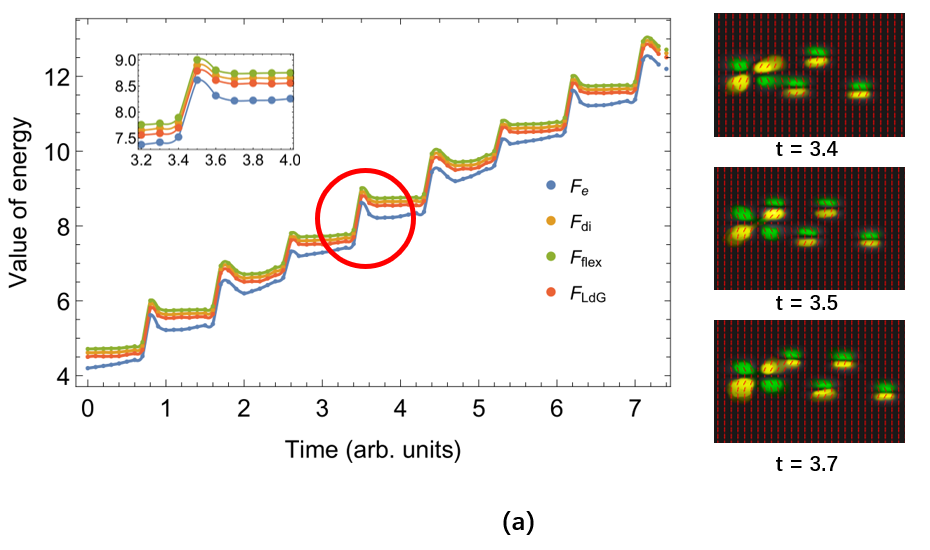}\\
		\includegraphics[width=0.65\textwidth]{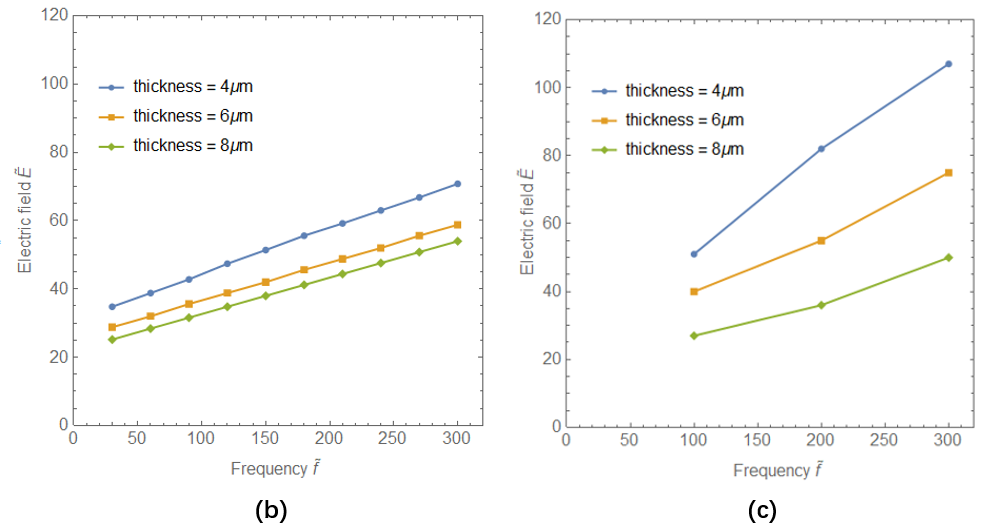}
		\caption{(a) Free energy of the system in the state corresponding to bullets emitted from a stable butterfly (orange band in FIG.~\ref{phasediagram}). The inset shows a zoomed-in portion of the free energy corresponding to the emission of an individual bullet. The panels on the right show director field configurations through cross polarizers corresponding to different times in the figure shown in the inset.  (b) Results of simulations for the threshold voltage for bullet emission from a butterfly for different sample thicknesses. (c) Experimental results for threshold voltage for different sample thicknesses.}
		\label{thicknessAndEnergy}
	\end{figure}

	It is of interest to examine the contributions of elastic energy $F_e$, the free energy of dielectric polarization $F_{di}$, and the flexoelectric polarization (equation ~(\ref{freeenergy})), as bullets are emitted from the wings of a stable butterfly. We show this process for $\widetilde{E_0} =38$ and $ \widetilde{\omega}=400$ in FIG.~\ref{thicknessAndEnergy}(a). These energies are set to zero in the uniform state (the director field is completely aligned along the $x-$axis). The elastic energy $F_e$ includes a combination of splay, bend, and twist contributions. Since the electric field is along the $z-$axis, the free energy of dielectric polarization $F_{di}$ should increase when more regions of the director field are aligned along the  $z-$axis. From the discussion of the flexoelectric free energy $F_{flex}$ in Section $\left(\uppercase\expandafter{\romannumeral2}\right)$, it is apparent that splay and bend deformations are key contributors. In FIG.~\ref{thicknessAndEnergy}(a), we show the data at the time when $E_z=E_0$, so that $F_{flex}$ is always positive. In the circle and the zoomed in panel in the figure, we focus on the peak of the free energy. The energies are small before $t=3.4$, where no bullet is emitted from the butterfly. They then increase ($t=3.4$ to $t=3.7$)  as a new bullet is generated from the upper branch of the butterfly. The peak reaches a maximum ($t=3.5$) when the bullet is about to be emitted from the butterfly, and it decreases when the bullet moves away. There are multiply peaks in the figure; each peak represents a new-born bullet emitted from a stationary butterfly.
	
	We also consider the effect of anchoring on soliton formation. FIG.~\ref{thicknessAndEnergy}(b) and (c) show the state in which the butterfly structure starts to emit bullets for channels of different thickness. We include results from simulations and experiments. Both sets of results show that, in a narrower channel, where anchoring has a stronger influence, a higher electric field is needed to emit bullets. We find that a linear relationship exists between the frequency and the electric field strength. When the voltage is large, the system can rearrange its structure in a shorter time interval. If the frequency is low, the material has sufficient time to undergo a complete structure change, and the solitary structure is broken. As a result, a higher voltage requires that a higher frequency be used if we want to generate a solitary structure. The inverse relationship between thickness and voltage is only observed over a small range of thickness.  For higher thickness, the effect of the confining surfaces is reduced, and solitons are no longer observed regardless of the applied voltage is. This is also seen in experiments~\cite{Li2018}.  

	\begin{figure} [H]
		\centering
	\includegraphics[width=0.55\textwidth]{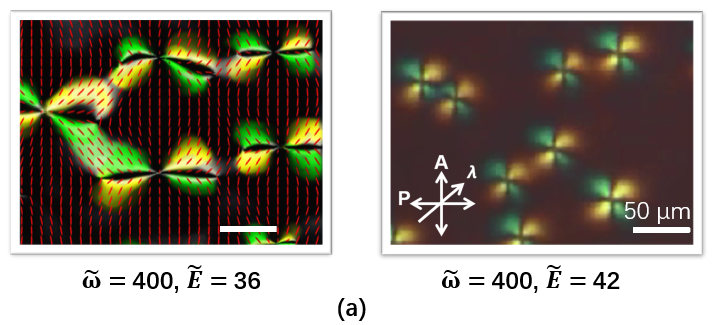}\\
	\includegraphics[width=0.55\textwidth]{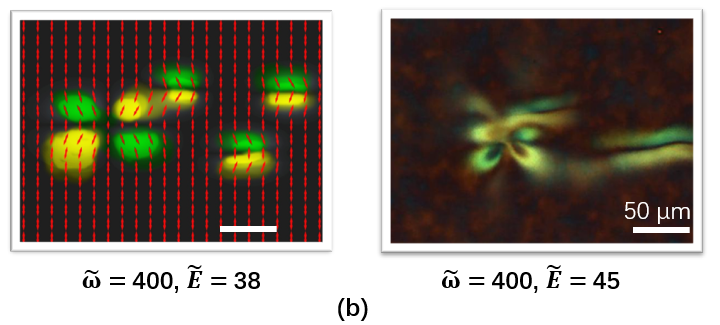}\\
	\includegraphics[width=0.55\textwidth]{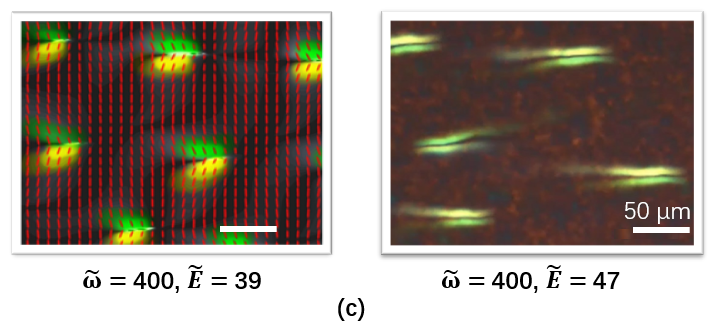}
	\caption{Comparison of results from simulations (left) and experiments (right) corresponding to the director field as observed through cross polarizers for nematic liquid crystal under different applied fields. In (a), several butterflies are observed. In (b), bullets are emitted from the butterfly. In (c), rapidly moving bullets are observed.}
	\label{compareWithExperiment}
	\end{figure}	
	
	FIG.~\ref{compareWithExperiment} focuses on the three most relevant regions of the phase diagram. We show results from simulations and experiments. When $\widetilde{E_0}=36$ and $\widetilde{\omega}=400$ (FIG.~\ref{compareWithExperiment} (a)), stable butterflies are observed and they move at very low speed. When the electric field reaches $\widetilde{E_0}=39$ at the same frequency (FIG.~\ref{compareWithExperiment} (c)), the butterflies shrink, and the system consists mostly of stripes and bullets. For electric fields between these two cases, e.g. $\widetilde{E_0}=38$ (FIG.~\ref{compareWithExperiment} (b)), we can observe that the bullets emerge from the butterfly created by the irregularity on the surface. The bullets may emanate from any wing of the butterfly. The bullets travel over a long distance in a direction perpendicular to the director. In experiments, the same structures are observed for the same frequency but at slightly higher voltages~\cite{das2022programming}. We attribute the discrepancy to the simplicity of our model and the fact that hydrodynamic effects have not been included in our model and calculations.
	
	\section{Conclusions}

In this work we have investigated the formation and propagation of solitons in a 3D nematic liquid crystal confined between to planar surfaces with planar anchoring. A minimal model of the nematic LC has been used to show that solitons are nucleated at a surface inhomogeneity with perpendicular anchoring, and are generated upon exposure to an AC field. A negative dielectric permittivity and flexoelectricity have been shown to be critical elements of the model required for soliton generation. The AC field causes the director field to adopt distinct states periodically and, for sufficiently high frequencies, the material is unable to relax to its stationary structure, thereby emitting solitons that travel at large speeds throughout the system. 
	
The model has been used to construct a state diagram as a function of the strength of the applied field and its frequency. Several regimes are predicted in simulations, including a uniform state, a stripe state, a chaotic state, and a soliton state, which only occurs over a narrow range of electric field strength. Good agreement is obtained between the results of simulations and our own experimental observations, serving to establish the validity of the model. Importantly, the findings presented in this work provide foundational knowledge with which to design new systems for the controlled production of solitons, thereby opening new opportunities for their use for applications in microfluidic transport, optical, and sensing technologies.

    \bibliographystyle{ieeetr}
	\bibliography{main.bib}

\end{document}